\documentclass[12pt]{article}
\usepackage{xcolor}
\usepackage{setspace}
\newcommand{\anon}{1}

\def\spacingset#1{\renewcommand{\baselinestretch}%
{#1}\small\normalsize} \spacingset{1}

\usepackage{appendix}
\usepackage{xr}
\usepackage{multirow}
\usepackage{threeparttable}
\usepackage{booktabs}
\usepackage[figuresright]{rotating}
\graphicspath{{figures/}}
\usepackage{natbib}
\usepackage{amsfonts,amsmath,mathrsfs,amssymb}

\DeclareMathOperator*{\argmax}{arg\,max}
\DeclareMathOperator*{\argmin}{arg\,min}
\usepackage{graphics}
\usepackage{graphicx}
\usepackage{subcaption}
\usepackage{ntheorem}
\usepackage[unicode]{hyperref}
\usepackage[noend]{algpseudocode}
\usepackage{algorithmicx,algorithm}
\allowdisplaybreaks[4]
\hypersetup{
    colorlinks=true,
    linkcolor=blue,
    citecolor=blue, 
}

\newtheorem{theorem}{Theorem}[section]

\newtheorem{corollary}{Corollary}[section]

\newtheorem{remark}{Remark}[section]

\begin{document}
\if1\anon
{
  \title{\bf Successive classification learning  for estimating quantile optimal treatment regimes}
 \author{Junwen Xia\(^{1,2}\), Jingxiao Zhang\(^{1,2*}\), and Dehan Kong\(^{3*}\)\\
{\small {\small {\it\(^1\)Center for Applied Statistics, Renmin University of China, Beijing, China}}}\\
{\small {\small {\it\(^2\)School of Statistics, Renmin University of China, Beijing, China }}}\\
{\small {\small {\it\(^3\)Department of Statistical Sciences, University of Toronto, Toronto, Ontario, Canada }}}\\
{\small {\small {\it\(^*\)email: zhjxiao@ruc.edu.cn; dehan.kong@utoronto.ca}}}}
  \maketitle
} \fi

\if0\anon
{
  \bigskip
  \bigskip
  \bigskip
  \begin{center}
    {\LARGE\bf Successive classification learning for estimating quantile optimal treatment regimes}
\end{center}
  \medskip
} \fi

\bigskip

\begin{abstract}
Quantile optimal treatment regimes (OTRs) aim to assign treatments that maximize a specified quantile of patients' outcomes. Compared to treatment regimes that target the mean outcomes, quantile OTRs offer fairer regimes when a lower quantile is selected, as it improves outcomes for vulnerable patients. In this paper, we propose a novel method for estimating quantile OTRs by reformulating the problem as a successive classification task, solvable via training a sequence of classifiers, each successive classifier built on the output of its predecessors. This reformulation enables us to leverage the powerful machine learning technique to enhance computational efficiency and handle complex decision boundaries. We also investigate the estimation of quantile OTRs when outcomes are discrete, a setting that has received limited attention in the literature. A key challenge is that direct extensions of existing methods to discrete outcomes often lead to inconsistency and ineffectiveness issues. To overcome this, we introduce a smoothing technique that maps discrete outcomes to continuous surrogates, enabling consistent and effective estimation. We provide theoretical guarantees to support our methodology, and demonstrate its superior performance through comprehensive simulation studies and real-data analysis. 
\end{abstract}

\noindent%
\vfill

\newpage
\spacingset{1.8} 

\section{Introduction}
Optimal treatment regimes (OTRs) aim to assign treatments based on patients' characteristics to achieve the best possible outcomes.
Previous work on estimating OTRs has primarily focused on optimizing the mean of the outcomes, commonly referred to as ``mean OTRs.'' Common methods include Q-learning \cite[]{watkins1992qlearning}, A-learning \cite[]{murphy2003optimal,shi2018highdimensional}, value search methods \cite[]{zhang2012a}, 
and classification methods \cite[]{zhao2012estimating,zhou2017residual,zhang2012estimating}. Despite the popularity of the mean criterion, other criteria such as the quantile may be more appropriate in many applications. For example, the empirical mean can be heavily influenced by extreme values. In such cases, the median is often preferred due to its robustness to outliers and heavy-tailed distributions.
{Moreover, in situations where outcomes are measured on an ordinal scale, such as cancer stages from 1 to 4, that support ranking but not arithmetic operations like averaging, the quantile criterion offers a more appropriate and interpretable measure.} OTRs that optimize a specific quantile, referred to as ``quantile OTRs'', can promote fairness, particularly when targeting lower quantiles \citep{wang2018quantileoptimal,fang2023fairness}. 
{Specifically, patients whose outcomes fall in the lower quantiles of the distribution are considered more vulnerable, as they experience relatively poorer results compared to the broader population. By explicitly targeting improvements in these lower quantiles, quantile OTRs aim to enhance outcomes for vulnerable patients and thereby promote fairness.}

With these insights, \cite{linn2017interactive} studied a general framework for estimating the quantile OTRs. Their method depends on model-based interactive Q-learning and therefore highly relies on a correctly specified model for the conditional distribution of the outcome given treatment and covariates. Later, \cite{wang2018quantileoptimal} studied the problem via value search methods, which established a consistent estimator of the counterfactual quantile given any regime and derived optimal regimes by maximizing it. {Their method can achieve the doubly robust property by leveraging both the propensity score model and the conditional quantile function of the outcome given treatment and covariates.} 
 Along this line, \cite{jiangDoublyRobustEstimation2017,zhouTransformationInvariantLearningOptimal2022} further studied quantile OTRs for survival data.

\cite{wang2018quantileoptimal} open the door to the application of the model-free method to quantile OTRs. However, the method exhibits several challenges. 
First, the optimization problem involved in the method is not concave, so there is no guarantee that the global maximizer can be derived even with the state-of-art genetic algorithm \cite[]{kramer2017genetic}. 
Second, the method primarily focuses on linear regimes. While linear regimes are interpretable, they may fail to capture the potentially complex and nonlinear structure of quantile OTRs in practice, thereby limiting the flexibility of the approach.
Third, when the outcomes are discrete, the method encounters issues of inconsistency and ineffectiveness. Specifically, the estimated quantile OTR may not consistently recover the true quantile OTR in terms of its corresponding quantile value across a wide range of quantile levels. Moreover, the expected counterfactual outcome under the estimated quantile OTR may not be the highest among all possible quantile OTRs. 
These challenges are discussed in detail in Section~\ref{sec:challenge}.

{To address these challenges, we propose a novel method called \textit{Successive Classification Learning (SCL)}}, which reformulates the problem of estimating quantile OTRs as a sequence of classification tasks solvable via convex optimization, thereby circumventing the non-convexity issue.
To accommodate nonlinear treatment regimes, we allow SCL to incorporate a Gaussian kernel, allowing for flexible and expressive decision boundaries. For discrete outcomes, we address inconsistency and ineffectiveness by introducing a smoothing step that maps discrete outcomes to continuous surrogates, allowing SCL to handle discrete and continuous outcomes within a unified framework. We summarize the main contributions of this work as follows.

First, we propose a novel method that leverages the classification learning to estimate quantile OTRs. The classification-based method has demonstrated notable strengths in estimating mean OTRs \cite[]{zhang2012estimating}, as they can harness machine learning techniques to enhance computational efficiency, avoid issues of local maximizers, and flexibly accommodate nonlinear treatment regimes. Motivated by these advantages, we seek to extend classification learning to the quantile setting. However, a direct extension is not feasible because the quantile function is not a linear operator. {To address this challenge, we recast the estimation of quantile OTRs as a successive classification problem, solvable by training a sequence of classification models, each one building upon the output of its predecessors and guided by a binary search procedure to refine the quantile OTR estimates.} {Leveraging this novel reformulation, we propose a doubly robust estimator for quantile OTRs by incorporating both the propensity score and the conditional survival function of the outcome given treatment and covariates.} 
{To enrich the applicability of our framework, we further extend our method to dynamic treatment settings, in which treatment decisions adapt over time according to patients’ intermediate responses to prior treatments, and to survival data, where outcomes of interest are subject to right censoring. These extensions are presented in Section S4 of the supplementary material.}

Second, we identify inconsistency and ineffectiveness issues when estimating quantile OTRs for discrete outcomes. 
To address these challenges, we introduce a smoothing technique that constructs continuous surrogates of the discrete outcomes. This allows us to define a smoothed quantile, and estimate quantile OTRs by maximizing this quantity. Importantly, although the smoothing alters the objective function to the smoothed quantile, we prove that the treatment regime maximizing the smoothed quantile also maximizes the original quantile. This theoretical guarantee justifies the use of smoothing for consistent and effective quantile OTR estimation in discrete settings. 

Third, we provide theoretical guarantees for our method. Our approach involves a sequence of interdependent classification tasks guided by a binary search procedure, posing significant analytical challenges due to the recursive dependence structure. To the best of our knowledge, asymptotic theory for such binary search-based procedures is largely unexplored. We address this gap by employing empirical process techniques. Our findings demonstrate that, when subjects with differing optimal treatments are well separated, the convergence rate of the quantile under the estimated regime to that under the true quantile OTR approaches $n^{-1/3}$ when employing a Gaussian kernel. Moreover, the convergence rate of our method is close to that of an oracle method, one possessing prior knowledge of the optimal quantile. En route to establishing these asymptotic results, we also derive novel theoretical insights for classification learning combined with the Gaussian kernel, thereby addressing gaps previously unfilled in \cite{zhang2012estimating}.


The remaining sections of this paper are organized as follows. Section \ref{method} presents the main methodology. Theoretical properties for SCL are provided in Section \ref{sec:consistency}. The finite sample performance is explored through simulations in Section \ref{simulations}. Section \ref{realdata} illustrates the application using the ACTG175 dataset. 
We end with a discussion in Section \ref{sec:discussion}. 

\section{Methodology} \label{method}
Let \(A\in\{0,1\}\) denote a binary treatment, \(\boldsymbol{X}\in \mathcal{X}\subseteq \mathbb{R}^d\) denote a \(d\)-dimensional baseline covariates, and \(Y\in\mathbb{R}\) denote the observed outcome of interest. We assume a larger value of \(Y\) indicating a better clinical outcome. Let \(Y(a)\) denote the counterfactual outcome under treatment \(A=a\in\{0,1\}\). An individual treatment regime (ITR) is a function map \(d: \mathcal{X}\to \{0,1\}\) from patients' baseline covariates space \(\mathcal{X}\) to the treatment space \(\{0,1\}\). Then, we define the counterfactual outcome under the ITR \(d\) as \(Y(d)=Y(0) I\{d(\boldsymbol{X})=0\} + Y(1) I\{d(\boldsymbol{X})=1\} \). We make the following assumption.

\vspace{0.3ex}
\textbf{Assumption 1:}
  (i) \(Y=Y(0)I(A=0)+Y(1)I(A=1)\); (ii) \(Y(0), Y(1) \perp \! \! \! \perp A | \boldsymbol{X}\); (iii) \(\epsilon<P(A=1|\boldsymbol{X})<1-\epsilon\) for some \(\epsilon>0\).

{Assumption 1 is a fundamental assumption in causal inference to ensure identification \cite[]{hernán2020causal}. Assumption 1(i) is referred to as the stable unit treatment value assumption (SUTVA), which requires each patient’s outcome depends solely on patient's own treatment. Assumption 1(ii) is referred to as the ignorability assumption, which guarantees that, conditional on \(\boldsymbol{X}\), treatment assignment $A$ is independent of the potential outcomes. Assumption 1(iii) is referred to as the positivity assumption, which means that every subgroup defined by covariates has a positive probability of receiving each treatment option.}


In this paper, our primary interest is the quantile OTRs \citep{wang2018quantileoptimal} defined as: 
\begin{equation}\label{Wang:quantile:objetive}
  d^*=\argmax_{d} Q_\tau\{Y(d)\},
\end{equation}
where \(\tau\) is the quantile level of the interest, and \(Q_\tau\{Y(d)\}\) is the \(\tau\)-th quantile of the outcome \(Y(d)\) defined as \( Q_\tau\{Y(d)\}=\sup\{q: S(q, d)> 1-\tau\}.\) Here, \(S(q,d)=P\{Y(d)> q\}\) is the counterfactual survival function under the regime \(d\) and it can be identified by
\begin{align*}
    S(q,d)=E\left[\frac{I(Y>q)I\{A=d(\boldsymbol{X})\}}{\pi^*(A|\boldsymbol{X})}\right],
\end{align*}
where \(\pi^*(a|\boldsymbol{x})=P(A=a|\boldsymbol{X}=\boldsymbol{x})\) is the propensity score. We utilize the survival function to define the quantile, as our novel method, SCL, is inspired by this definition, and such a quantile definition is equivalent to the traditional definition using the cumulative distribution function. See Section S5.1 of the supplementary material for details.

\subsection{Challenges of the value search method} \label{sec:challenge}
Suppose the observed data \(\{\boldsymbol{X}_i, Y_i,A_i\}, i=1, \ldots, n\) are independent and identically distributed copies of \(\{ \boldsymbol{X}, Y, A\}\). 
To estimate the quantile OTRs, \cite{wang2018quantileoptimal} first estimate \(Q_\tau\{Y(d)\}\) by
\begin{equation} \label{wang.quantile}
  \widehat{Q}_\tau\{Y(d)\}=\underset{q}{\argmin } \frac{1}{n} \sum_{i=1}^{n} \frac{I\{d(\boldsymbol{X}_i)=A_i\} }{\pi^*(A_i| \boldsymbol{X}_i)} \rho_{\tau}\left(Y_{i}-q\right),
\end{equation}
where \(\rho_{\tau}\left(u\right)= u\{\tau-I(u<0)\} \) is the check loss. Then they consider the set of linear ITRs \(\mathcal{D}=\{d(\boldsymbol{x})=I(\beta_0+{\boldsymbol{x}}^T\boldsymbol{\beta}> 0): |\beta_0|=0,1\text{ and } \boldsymbol{\beta} \in \mathbb{R}^{d}\}\), and estimate quantile OTRs by maximizing \(\widehat{Q}_\tau\{Y(d)\}\) over the set \(\mathcal{D}\) based on the genetic algorithm  \citep{kramer2017genetic}. 


The value search method proposed by \cite{wang2018quantileoptimal} lays a strong foundation for the estimation of quantile OTRs. However, it exhibits several challenges:
\begin{enumerate}
\item Non-convexity: The objective function $\widehat{Q}_\tau\{Y(d)\}$ is not concave with respect to the parameters \((\beta_0,\boldsymbol{\beta}^T)^T\). As a result, the solution based on the genetic algorithm may correspond to a local maximizer, potentially leading to a harmful treatment regime, and posing significant risks in real-world applications. This issue becomes more critical when \(\boldsymbol{\beta}\) becomes high dimensional \citep{shi2018forward}. We provide a simulation in Section S1.1 of the supplementary material. In that particular setting, \cite{wang2018quantileoptimal}'s method fails to find the global maximizer \(61.5\)\% of the time.

\item Regime linearity: \cite{wang2018quantileoptimal} mainly studied the linear regime. While the linear regime is easy to interpret, in practice, there may be greater interest in more flexible, nonlinear quantile OTRs. For example, consider a scenario where the outcome model follows a simple heteroscedastic linear form, i.e., \(Y=m(\boldsymbol{X})+A g(\boldsymbol{X}) + h(\boldsymbol{X}, A)\epsilon \), where \(\epsilon\) is a random noise and \(m(\boldsymbol{X})\), \(g(\boldsymbol{X})\), and \(h(\boldsymbol{X}, A)\) are all linear functions. In this case, the mean OTRs are linear, however, the quantile OTRs are nonlinear. A detailed derivation of this result is provided in Section S1.2 of the supplementary material. This highlights that even simple linear outcome models can yield nonlinear quantile OTRs, emphasizing the importance of exploring nonlinear approaches for quantile OTRs. 

\item Inconsistency: In the OTR literature, value consistency is a commonly established property \citep{zhao2015doubly}. For continuous outcomes, \citet{wang2018quantileoptimal}'s method enjoys quantile value consistency; i.e., $Q_{\tau}\{Y(\hat{d})\}$ converges in probability to {$\max_{d\in\mathcal{D}}$\\$Q_{\tau}\{Y(d)\}$}. However, when their method is applied directly to discrete outcomes, this consistency property no longer holds. We demonstrate this through a toy simulation study in Section S1.3 of the supplementary material. We compute the empirical mean squared error (MSE) obtained by \citet{wang2018quantileoptimal}'s method, defined as the average of $[Q_{\tau}\{Y(\hat{d})\}-\max_{d\in\mathcal{D}}Q_{\tau}\{Y(d)\}]^2$ across $200$ Monte Carlo replications. The results are presented in Figure \ref{fig:inconsitency}, which shows that the empirical MSE of \cite{wang2018quantileoptimal}'s method does not diminish as the sample size increases across various quantile levels. In contrast, our proposed method exhibits a decreasing MSE when sample size increases, providing evidence of consistency. Notably, the MSE of our method can reach zero due to the discrete nature of the outcome. Full simulation details and an extended explanation of these results are provided in Section S1.3 of the supplementary material. The inconsistency issue stems from the fact that the quantile estimator (\ref{wang.quantile}) is not a consistent estimator of the quantile \(Q_{\tau}\{Y(d)\}\) when the outcome is discrete \cite[]{machado2005quantiles}. As a consequence, maximizing (\ref{wang.quantile}) does not necessarily yield a consistent estimator of the quantile OTRs.

\begin{figure}
  \centering
  \includegraphics[width=0.95\textwidth]{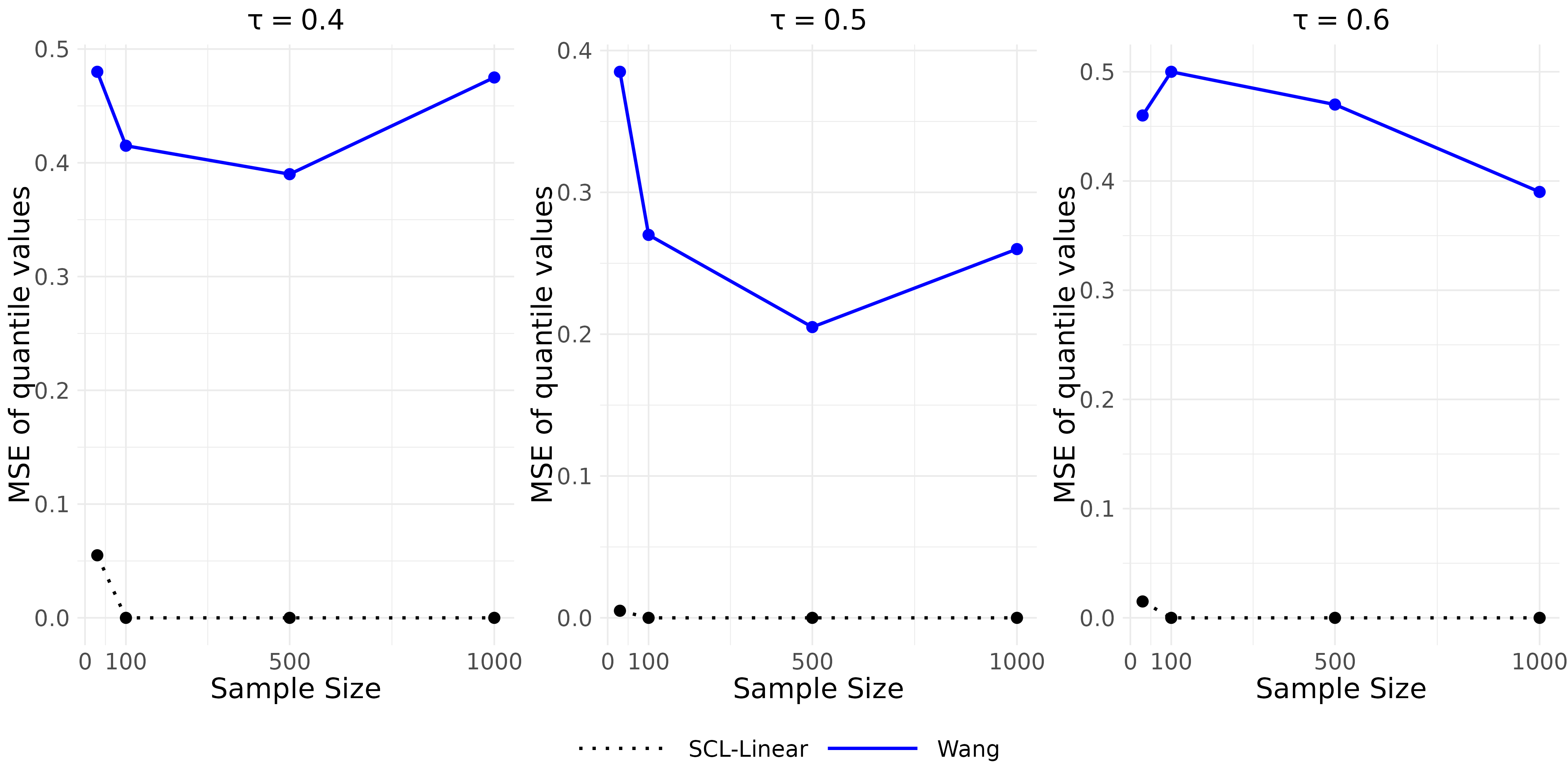}
  \caption{MSE of the quantile values given the estimated regimes based on 200 repeated simulations. ``SCL-Linear'' denotes our proposed method with the linear kernel. ``Wang'' denotes the method by maximizing (\ref{wang.quantile}).}
  \label{fig:inconsitency}
\end{figure}

\item {Ineffectiveness: Define \(\mathcal{D}^*=\{\tilde{d}\in\mathcal{D}:Q_{\tau}\{Y(\tilde{d})\}=\max_{d\in\mathcal{D}}Q_{\tau}\{Y({d})\}\}\) as the set of quantile OTRs. Since \(\mathcal{D}^*\) typically comprises multiple regimes in the case of discrete outcomes, it may be desirable to identify the one that best satisfies an additional criterion. In this paper, we introduce a natural criterion that compares their expected counterfactual outcomes, \(E\{Y(d^*)\}\) for each \(d^* \in \mathcal{D}^*\), and selects the regime with the highest expected counterfactual outcomes. We refer to such a regime as an \textit{effective} quantile OTR. However, the quantile OTR obtained by solving \eqref{Wang:quantile:objetive} \citep{wang2018quantileoptimal} is not guaranteed to be effective under this criterion. 
Therefore, estimating quantile OTRs via maximizing the empirical version of $Q_{\tau}\{Y(d)\}$, i.e., \(\hat{d}= \argmax_{d\in\mathcal{D}} \widehat{Q}_{\tau}\{Y(d)\}\), also suffers from the ineffectiveness issue.
We have provided details in Section S1.4 of the supplementary material.}
\end{enumerate}

\subsection{Successive classification learning for continuous outcomes} \label{cont}
To address these challenges, we propose a novel approach inspired by classification learning. Classification learning methods have demonstrated notable strengths in estimating mean OTRs \cite[]{zhang2012estimating, zhao2012estimating}; for example, they can leverage machine learning techniques to accelerate computation, avoid issues of local maximizers, and accommodate flexible, nonlinear regimes. However, they cannot be directly extended to quantile OTRs because the quantile function is not a linear operator. 
{To overcome this limitation, we recast the estimation of quantile OTRs as a successive classification problem, solvable through training a sequence of classifiers, each building on the outputs of its predecessors.} We refer to our method as \textit{Successive Classification Learning (SCL)}.

To illustrate our idea, we first consider the continuous counterfactual outcome \(Y(d)\) with a strictly decreasing survival function on its support \([a, b]\), where we can allow \(a\) and \(b\) to be infinity. Since for a continuous and strictly decreasing function \(H(x)\) defined on \(x\in[a,b]\) with \(H(a)=1\) and \(H(b)=0\), \(\sup\{x:H(x)>1-\tau\}=\max\{x:H(x)\geq 1-\tau\}\) for \(\tau\in(0,1)\). 
We then have \(Q_\tau\{Y(d)\} = \sup\{q: S(q, d) > 1 - \tau\}= \max\{q: S(q, d) \geq 1 - \tau\}\). The quantile OTRs satisfy \(d^*=\argmax_d Q_{\tau}\{Y(d)\}= \argmax_d\max\{q:S(q,d)\geq 1-\tau\}\). We can then obtain the quantile OTRs by solving the following optimization problem
    \begin{align*}
  d^*=\underset{d}{\argmax} ~ q~,
  ~~ {\rm subject~to~} S(q, d) \geq 1-\tau.
 \end{align*}


By this neat reformulation, the quantile OTRs can be solved as a constrained optimization problem by varying \(q\in[a,b]\) from large to small and checking whether each \(q\) satisfies the constraint \(\max_{d} S(q, d) \geq 1 - \tau\). Denote the largest \(q\) satisfying \(\max_{d} S(q, d) \geq 1 - \tau\) by \(q^*\). Then \(d_{q^*}^* = \argmax_{d} S(q^*, d)\) is the solution to the constrained optimization problem and thus the quantile OTRs. Furthermore, since the counterfactual quantile given the quantile OTRs is \(q^*\), i.e., \(Q_{\tau}\{Y(d_{q^*}^*)\}=\max\{q: S(q,d_{q^*}^*)\geq 1-\tau\}=q^*\), it follows that \(q^*\) is the optimal quantile \(\max_d Q_{\tau}\{Y(d)\}\).

Varying $q$ from largest to smallest and searching for the solution is computationally expensive. Instead, we employ a binary search method for efficient computation, leveraging the nice property that \(\max_dS(q, d)\) is a continuous and strictly decreasing function of \(q\). Specifically, we begin our search within a large enough bounded interval \((l, r)\) that contains the optimal quantile \(q^*\). Set \(m = (l + r)/2\) and compare \(\max_dS(m, d)\) with \(1 - \tau\). If they are equal, we consider that the optimal quantile \(q^*\) is found and \(d_{q^*}^*=\argmax_d S(q^*,d)\) is the quantile OTRs. Otherwise, if \(\max_dS(m, d) > 1 - \tau\), we continue our search to the interval \((m, r)\); else, we proceed within the interval \((l, m)\). 


Building on the preceding discussion, the estimation of quantile OTRs $d_{q^*}^*$ can be carried out by first estimating the optimal quantile $q^*$, and then using this value to construct an estimate of $d_{q^*}^*$.
The estimation of $q^*$ can be obtained via the previously described binary search procedure, wherein the optimal survival functions $\max_d S(q, d)$ are replaced by their empirical estimates. Since the survival function can be expressed as an expectation $S(q, d)=E[I\{Y(d)>q\}]$, we can borrow the idea of classification learning \citep{zhang2012estimating} to estimate $\max_{d} S(q, d)$ and ultimately $d_{q^*}^* = \arg\max_d S(q^*, d)$, taking advantage of the favorable optimization and statistical properties inherent to classification learning.
Overall, the estimation of quantile OTRs is effectively recast as a successive classification learning problem.

We now explicitly detail the estimation procedure, beginning with the estimation of $\max_d S(q, d)$ for any fixed $q$. Our strategy involves estimating $d_q^*=\argmax_dS(q,d)$ with classification learning and plugging this estimate into an appropriate estimator for $S(q, d)$.

A classification learning framework with the doubly robust property is adopted for $d_q^*$ \citep{zhang2012estimating}, which is underpinned by the identification of $S(q, d)$ \cite[]{robins1994estimation}:
\begin{equation} \label{eq:doublysurvival}
    {S}(q, d)=E\left[\psi(1,q; {\pi}^*, {g}^*)d(\boldsymbol{X})+\psi(0,q; {\pi}^*, {g}^*)\{1-d(\boldsymbol{X})\}\right],
\end{equation}
where
\begin{align*}
     \psi(a, q; \pi^*, g^*) = \frac{I(Y > q) - g^*(q; \boldsymbol{X}, a)}{\pi^*(a | \boldsymbol{X})} I(A = a) + g^*(q; \boldsymbol{X}, a).
\end{align*}
Here, $\pi^*(a | \boldsymbol{X})$ denotes the propensity score, and $g^*(q; \boldsymbol{X}, a) = P(Y > q | \boldsymbol{X}, A = a)$ represents the conditional survival function.

The doubly robust classification idea is to reformulate the \({d}_q^*\) as a solution to the weighted classification problem \cite[]{zhang2012estimating}:
\begin{equation}
  \label{eq:itrs}{d}_q^*=\argmax_{d} {S}(q,d)
  =\argmin_{d}E [|{C}(\boldsymbol{X},q; {\pi}^*,{g}^*)|I\{Z^* \neq   d(\boldsymbol{X})\}],
\end{equation}
where \(C(\boldsymbol{X},q;\pi^*,{g}^*)={\psi}(1,q; {\pi}^*, {g}^*)-{\psi}(0,q; {\pi}^*, {g}^*)\) and \(Z^*=I\{C(\boldsymbol{X},q;\pi^*,{g}^*)>0\}\).

Note that regimes $d(\boldsymbol{X})$ can always be represented as $I\{f(\boldsymbol{X}) > 0\}$ for some decision function $f$. The classification approach then employs the convex hinge loss $\phi\{(2Z^* - 1)f(\boldsymbol{X})\}$ as a surrogate for the non-convex 0-1 loss $I\{Z^* \neq d(\boldsymbol{X})\}$ \cite[]{zhao2012estimating}, where the hinge loss function is defined as $\phi(x) = (1 - x)_+ = \max\{1 - x, 0\}, \, x \in \mathbb{R}$.
{Replacing the 0-1 loss with the hinge loss offers several advantages. First, the convexity of the hinge loss enables efficient optimization. Second, the hinge loss $\phi\{(2Z^* - 1)f(\boldsymbol{X})\}=\{1-(2Z^* - 1)f(\boldsymbol{X})\}_{+}$ remains positive not only when $\boldsymbol{X}$ is misclassified, but also when \(\boldsymbol{X}\) is correctly classified yet lies close to the decision boundary \cite[]{cortes1995support}.
This property encourages the samples $\{\boldsymbol{X}_i\}_{i=1}^n$ to be pushed farther away from the decision boundary, thereby enhancing both robustness and generalization of the estimated regime.}

With the hinge loss function, we can estimate \(d_q^*(\boldsymbol{x})\) by \(\hat{d}_q(\boldsymbol{x})=I\{\hat{f}_q(\boldsymbol{x})>0\}\), where
\begin{equation} \label{surrogate}
  \hat{f}_q=\argmin_{f\in\mathcal{F}}\frac{1}{n}\sum_{i=1}^n |C(\boldsymbol{X}_i,q; \hat{\pi},\hat{g})|\phi\{(2\widehat{Z}_i-1)f(\boldsymbol{X}_i)\}+\lambda_n\|f\|^2.
\end{equation}
Here, \(\hat{\pi}\) is an estimator for the propensity score \(\pi^*\), and \(\hat{g}\) is an estimator for the conditional survival function \(g^*\), where the estimation of these nuisance functions is discussed in Section \ref{nuisance}. Additionally, \(\widehat{Z}_i=I\{{C}(\boldsymbol{X}_i,q; \hat{\pi},\hat{g})>0\}\) is the empirical analogue of \(Z_i^*\), constructed by replacing the nuisance functions by their estimators. The regularity term \(\lambda_n\|f\|^2\) is added to avoid overfitting with \(\lambda_n\) as the tuning parameter. We also consider a prespecified set of decision functions \(\mathcal{F}=\{b+h(\boldsymbol{x}):b\in\mathbb{R},h(\boldsymbol{x})\in\mathcal{H}_{k}\}\) and the norm \(\|f\|^2=\|h\|_k^2\),  where \(\mathcal{H}_{k}\) denotes a reproducing kernel Hilbert space (RKHS) and \(\|h\|_k\) is the norm induced by the inner product of the RKHS. To derive linear quantile OTRs, we consider \(\mathcal{H}_k\) as the RKHS reproduced by a linear kernel \(\left\langle \boldsymbol{x}, \boldsymbol{z} \right\rangle \), where \(\boldsymbol{x},\boldsymbol{z}\in\mathcal{X}\). For nonlinear quantile OTRs, we adopt the Gaussian kernel, 
\(
  k_{\sigma_n}(\boldsymbol{x}, \boldsymbol{z})=\exp \left(-\sigma_n^{2}\|\boldsymbol{x}-\boldsymbol{z}\|^{2}\right),
\) where \(\boldsymbol{x},\boldsymbol{z}\in\mathcal{X}\) and the parameter \(\sigma_n\) controls the complexity of the function reproduced by the kernel. 
The parameters \(\lambda_n\) (and \(\sigma_n\)) can be chosen via 10-fold cross-validation. For further details on solving the minimization problem (\ref{surrogate}), we refer the reader to \citet{zhao2012estimating}.

Building on the identification (\ref{eq:doublysurvival}), we adopt the following estimator for $\max_dS(q, d)=S(q, d_q^*)$ \cite[]{jiangDoublyRobustEstimation2017}: 
\begin{align} \label{eq:smoothedsurvival}
    \widehat{S}_{h_n}(q,\hat{f}_q)=\frac{1}{n}\sum_{i=1}^n
    \left(\psi_i(1,q; \hat{\pi}, \hat{g})\Phi\{\hat{f}_q(\boldsymbol{X}_i)/h_n\}+\psi_i(0,q; \hat{\pi}, \hat{g})[1-\Phi\{\hat{f}_q(\boldsymbol{X}_i)/h_n\}]\right),
\end{align}
where \( \Phi(x),x\in\mathbb{R}, \) denotes the cumulative distribution function of the standard normal distribution, and \( h_n \) serves as a bandwidth parameter. This estimator employs a soft estimated regime \(\Phi\{\hat{f}_q(\boldsymbol{x}) / h_n\}\) rather than the hard estimated regime \(\hat{d}_q(\boldsymbol{x})= I\{\hat{f}_q(\boldsymbol{x})> 0\}\) {to facilitate the theoretical development.} For the choice of the bandwidth parameter \(h_n\), we recommend \(h_n=0.2/\log n\), which performs well in our simulation and satisfies the conditions to establish the consistency result in Theorem \ref{theorem3}.

\begin{remark}
    {The soft regime $\Phi\{\hat{f}_q(\boldsymbol{x})/h_n\}$ to construct $\widehat{S}_{h_n}(q,\hat{f}_q)$ in (\ref{eq:smoothedsurvival}) may be replaced by the hard regime $I\{\hat{f}_q(\boldsymbol{x})>0\}$, although no theoretical guarantees exist as the indicator function $I\{\hat{f}_q(\boldsymbol{x})>0\}$ possesses an inherent jump discontinuity. A simulation provided in the Section S3.4 of the supplementary material reveals that these two regimes exhibit similar performance. Thus, we favor the soft regime for its theoretical guarantee.}
\end{remark}

The optimal quantile estimate $\hat{q} $ for \(q^*\) is then obtained by the binary search procedure specified previously by replacing \(\max_dS(q,d)\) with its estimate \(\widehat{S}_{h_n}(q,\hat{f}_q)\). Once $\hat{q}$ is determined, given that \(\hat{d}_q(\boldsymbol{x})\) is an estimator of \(d_q^*(\boldsymbol{x})\) for all \(q\), we can obtain the estimator $\hat{d}_{\hat{q}}(\boldsymbol{x}) = I\{\hat{f}_{\hat{q}}(\boldsymbol{x}) > 0\}$ for quantile OTRs \(d_{q^*}^*(\boldsymbol{x})\).

We have summarized our successive classification learning procedure in Algorithm \ref{alg_1}. There are several tuning parameters \( l^1 \), \( r^1 \), \( \kappa_n \), and \( \epsilon_n \) that need to be properly selected for the binary search procedure. The \(l^1\) and \(r^1\) should ensure that \(q^*\) is included in the interval \((l^1,r^1)\). If the support of \(Y\) is a bounded interval, we can choose \((l^1,r^1)\) that encompasses the support of \(Y\). Otherwise, we can empirically choose \(l^1\) and \(r^1\) as the minimum and the maximum of the observed outcome, respectively. The \(\kappa_n\) and \(\epsilon_n\) should be small to avoid introducing a large bias when estimating the optimal quantile \(q^*\). We recommend \(\kappa_n=6^{-1}n^{-1/2}\operatorname{sd}(Y)\) and \(\epsilon_n=0.5n^{-1/2}\), where \(\operatorname{sd}(Y) \) is the sample standard deviation of random variable \( Y \). It performs well in our simulation and satisfies the conditions to establish the consistent result in Theorem \ref{theorem3}.

\begin{algorithm}[t] 
  \caption{The algorithm for estimating the quantile OTRs.}\label{alg_1}
  \hspace*{0.02in} {\bf Input:} Input the tolerance errors \(\epsilon_n, \kappa_n \), the range of the feasible value \((l^1,r^1)\), the quantile level \(\tau\), and the observed data \(\{\boldsymbol{X}_i,Y_i,A_i\}, i=1, 2, ..., n\).\\
  \hspace*{0.02in} {\bf Output:} Output the estimated quantile OTRs \(\hat{d}\).
  \begin{algorithmic}
  \State 1) Estimate the nuisance functions \(\pi^*\) and \(g^*\), denoted by \(\hat{\pi}\) and \(\hat{g}\).
  \vspace{0.2em}
  \vspace{0.2em}
  \State 2) For \(b=1,2,...,\) repeat steps (2.1)-(2.6) until \(r^b-l^b\leq \kappa_n\).
  \vspace{0.2em}
  \State \quad (2.1) Set \(m^b=(l^b+r^b)/2\). 
  \vspace{0.2em}
  \State \quad (2.2) For a series of tuning parameters \(\lambda_n\) (and \(\sigma_n\) if we choose a Gaussian kernel), choose the best parameter via 10-fold cross-validation. 
  \vspace{0.2em}
  \State \quad (2.3) With the best parameter, we solve
  \begin{align*}
      \hat{f}_{m^b}={\argmin}_{f\in \mathcal{F}}\frac{1}{n}\sum_{i=1}^n |{C}(\boldsymbol{X}_i,m^b;\hat{\pi},\hat{g})|\phi\{(2\widehat{Z}_i-1)f(\boldsymbol{X}_i)\}+\lambda_n\|f\|^2.
  \end{align*}

  \State \quad (2.4) \(\mathbf{If}\) \(|\widehat{S}_{h_n}(m^b,\hat{f}_{m^b})-1+\tau| \leq \epsilon_n\) 
  \vspace{0.2em}
  \State \quad \quad \quad \(\mathbf{Break}\).
  \vspace{0.2em}
  \State \quad (2.5) \(\mathbf{Else}\) \(\mathbf{if}\) \(\widehat{S}_{h_n}(m^b, \hat{f}_{m^b})\geq 1-\tau\)
  \vspace{0.2em}
  \State \quad \quad \quad Update \(l^{b+1}=m^{b}\), \(r^{b+1}=r^{b}\).
  \vspace{0.2em}
  \State \quad (2.6) \(\mathbf{Else}\) 
  \vspace{0.2em}
  \State \quad \quad \quad Update \(l^{b+1}=l^{b}\), \(r^{b+1}=m^{b}\).
  \vspace{0.2em}
  \State 3) Output \(\hat{d}(\boldsymbol{x})=I\{\hat{f}_{\hat{q}}(\boldsymbol{x})> 0\}\), where \(\hat{q}=m^B\) and \(B\) is the number of iterations.
  \end{algorithmic}
  \end{algorithm}  

  \begin{remark}
     In the preceding discussion, we assumed that the survival function of $Y(d)$ is strictly decreasing and continuous over its support $[a,b]$. This assumption can be relaxed to require only that the survival function $S(q,d)$ is strictly decreasing and continuous in $q$ within a neighborhood of $\max_d Q_{\tau}\{Y(d)\}$. 
  \end{remark}

\subsection{Successive classification learning for discrete outcomes} \label{discrete}
\begin{figure}
  \centering
  \includegraphics[width=0.65\textwidth]{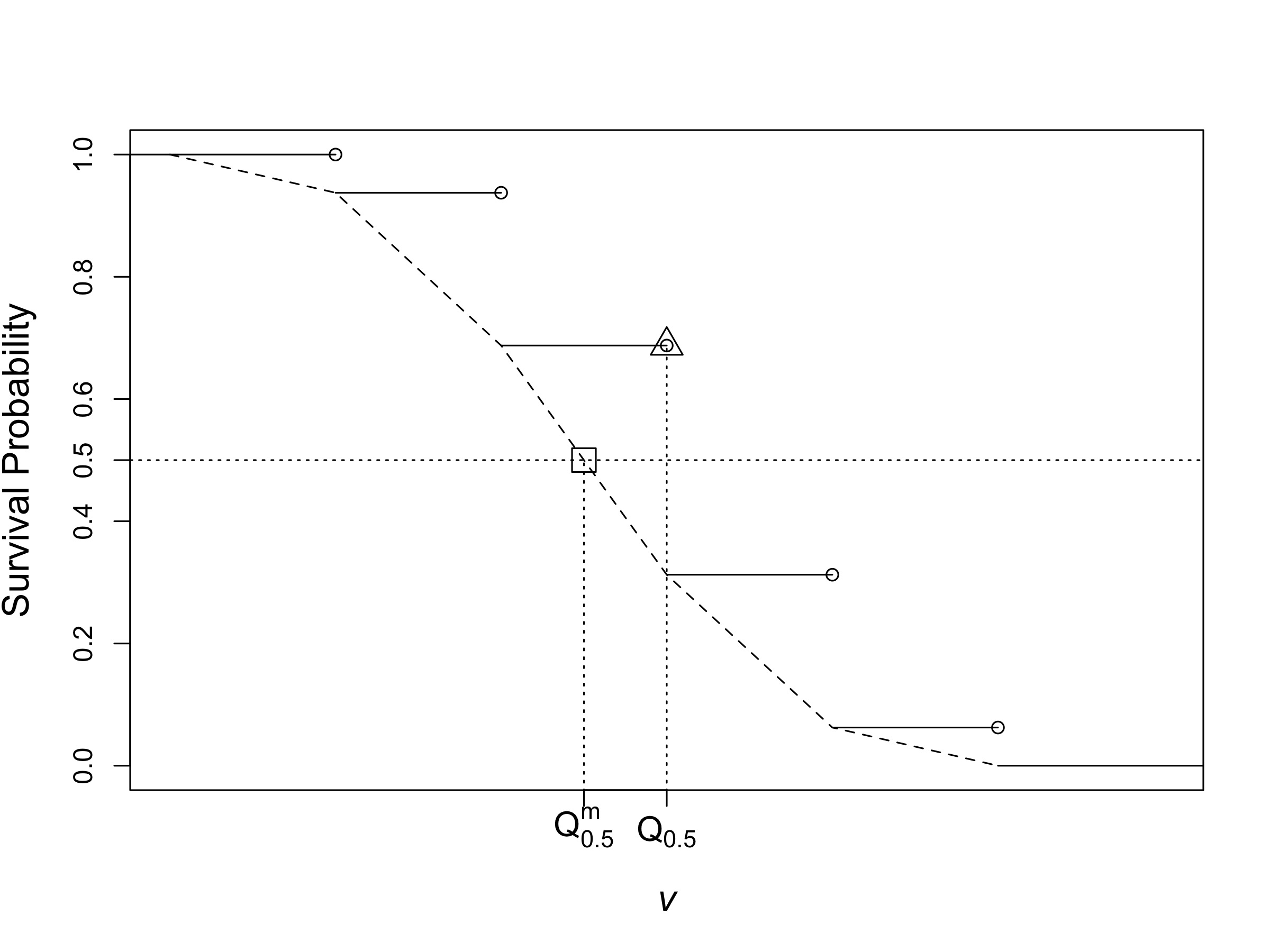}
  \caption{An illustration of the smoothed survival function and quantiles for discrete outcomes. The solid line represents the original survival function $S(q,d)$, with circles indicating all possible values that 
 $ Y(d)$ can take. The dashed line depicts the smoothed survival function. The dotted horizontal lines represent the quantile level $ \tau=0.5$, highlighting the difference in median definitions: the rectangle denotes the smoothed median, while the triangle represents the original median.}
  \label{mid_survival}
\end{figure}

We next consider the scenario when the outcome of interest is discrete. As we mentioned in Section \ref{sec:challenge}, discrete outcomes bring additional challenges due to inconsistency and ineffectiveness issues. This is primarily because the survival function $S(q,d)$ is piecewise constant with respect to $q$, which implies that the quantile $Q_{\tau}\{Y(d)\} $ is also piecewise constant with respect to the quantile level \(\tau\). To overcome the challenges, we propose to smooth the survival function $S(q,d)$. For convenience, suppose integers \(v_1<v_2<...<v_l\) are the possible distinct values of the counterfactual outcome \(Y(d)\), and let \(v_0=v_1-1\). Here, we allow \(l\) to be infinity. We then define the smoothed version of $S(q,d)$ as
\begin{equation} \label{smoothed_survival}
  S^{m}(q,d)=\left\{\begin{array}{ll}
    1 & \text { if } q\leq v_0, \\
    S(v_{k},d) & \text { if } q=v_{k}, k=1, 2, \ldots, l-1, \\
    \lambda S(v_{k},d)+(1-\lambda) S(v_{k+1},d) & \text { if } q=\lambda v_{k}+(1-\lambda) v_{k+1}, 0<\lambda<1, \\
    & k=0, 1, \ldots, l-1, \\
    0 & \text { if } q\geq v_{l}.
    \end{array}\right.
\end{equation}
Subsequently, we can define the smoothed quantile function 
\begin{equation} \label{smoothed_quantile}
  Q_{\tau}^{m}\{Y(d)\}=\sup \{q: S^{m}(q, d)> 1-\tau\}. 
\end{equation}
Figure \ref{mid_survival} gives an example of a smoothed survival function and the corresponding smoothed quantile. 

The key idea of our approach is to map the discrete outcome to a continuous outcome by uniformly distributing the mass at $ v_j $ over the interval $ [v_{j-1}, v_{j}]$. 
Define the set of regimes maximizing the smoothed quantile by \(\tilde{\mathcal{D}}^* = \{\tilde{d}: Q_{\tau}^m\{Y(\tilde{d})\} = \max_{d} Q_{\tau}^m\{Y(d)\} \}\) and recall the set of quantile OTRs \(\mathcal{D}^* = \{\tilde{d}: Q_{\tau}\{Y(\tilde{d})\} = \max_{d} Q_{\tau}\{Y(d)\} \}\).  Surprisingly, we can prove that $ \tilde{\mathcal{D}}^* \subseteq \mathcal{D}^*$ (see Theorem S2 in Section S5.3 of the supplementary material). Therefore, the regime $\tilde{d}^*\in \tilde{\mathcal{D}}^* $ belongs to the solution set \(\mathcal{D}^*\). 
Then, since $S^m(q,d)$ is a continuous function of $q$, we can {consistently} estimate the quantile OTRs using Algorithm \ref{alg_1}, substituting the estimated optimal decision functions $\hat{f}_q$ and the estimated optimal survival functions $\widehat{S}_{h_n}(q,\hat{f}_q)$ with their smoothed counterparts, $\hat{f}_q^m$ and $\widehat{S}_{h_n}^m(q,\hat{f}_q^m)$, obtained through survival smoothing. 

In particular, we smooth $S(q,d)$ given in (\ref{eq:doublysurvival})  using the procedure in (\ref{smoothed_survival}).  
This yields 
\begin{align*}
    {S}^m(q, d)=E\left[\psi^m(1,q; {\pi}^*, {g}^*)d(\boldsymbol{X})+\psi^m(0,q; {\pi}^*, {g}^*)\{1-d(\boldsymbol{X})\}\right],
\end{align*}
where \(\psi^m(a,q; {\pi}^*, {g}^*)\) is the smoothed counterpart of \(\psi(a,q; {\pi}^*, {g}^*)\). The explicit expression for \(\psi^m(a,q; {\pi}^*, {g}^*)\) is given in (3) of the supplementary material.

Then, we apply the classification idea to derive $\hat{f}_q^m$ and subsequently construct $\widehat{S}_{h_n}^m(q,\hat{f}_q^m)$ following a similar derivation in Section \ref{cont}. We have provided detailed procedures to compute $\hat{f}_q^m$ and $\widehat{S}_{h_n}^m(q,\hat{f}_q^m)$ in Section S2 of the supplementary material. Additionally, a detailed algorithm tailored for discrete outcomes is presented in Algorithm S1 of the supplementary material.

A key advantage of focusing on \(\tilde{\mathcal{D}}^*\) from the solution set \(\mathcal{D}^*\) is that it can exclude quantile OTRs that are close to the non-optimal regime and therefore improve the performance when estimating quantile OTRs. When the homoscedastic treatment effect assumption (Assumption 9 in Section \ref{sec:resultdiscrete}) holds, only the effective quantile OTRs remain in the set \(\tilde{\mathcal{D}}^*\) (see Theorem S3 in Section S5.3 of the supplementary material), which helps solve the ineffectiveness issue.

In summary, by applying a smoothing technique to 
$ S(q,d)$, our proposed estimator based on the smoothed survival function $S^{m}(q,d)$ achieves consistency for the quantile value and demonstrates strong performance. The formal theoretical results are presented in Section~\ref{sec:resultdiscrete}.

\begin{remark} \label{remark:alternative}
     {Alternative smoothing techniques beyond (\ref{smoothed_survival}) can be employed provided that the resulting smoothed survival function \(S^{m}(q,d)\) is continuous, strictly decreasing, and matches the original survival function \(S(q,d)\) at all points \(q\in\{v_1,\dots,v_l\}\). Let \(\tilde{\mathcal{D}}_1^*\) and \(\tilde{\mathcal{D}}_2^*\) denote the sets of regimes that maximize the smoothed quantile under two distinct smoothing techniques satisfying the above conditions. We can show that \(\tilde{\mathcal{D}}_1^* = \tilde{\mathcal{D}}_2^* \subseteq \mathcal{D}^*\), demonstrating that different smoothing techniques lead to the same quantile OTRs. More details about the alternative smoothing techniques can be found in Section 5.4 of the supplementary material. We also implement one of the alternative smoothing method and provide a simulation comparing it with (\ref{smoothed_survival}) in Section 3.5 of the supplementary material. The results reveal no notable differences between the two methods. Therefore, we favor the smoothed survival function in (\ref{smoothed_survival}), since it is simper and computationally easier.}
\end{remark}

\subsection{Estimation of the nuisance functions} \label{nuisance}
Our procedure involves estimating the nuisance functions: the propensity score \(\pi^*\) and the conditional survival function \(g^*\). In our paper, we consider the logistic regression for the propensity score model, i.e.,   $\pi(A=1|\boldsymbol{X})=\mathrm{expit}(\alpha_0+\boldsymbol{X}^T\boldsymbol{\alpha})$, where \(\mathrm{expit}(x)=\mathrm{exp}(x)/\{1+\mathrm{exp}(x)\}\), \(\alpha_0\in\mathbb{R}\), and \(\boldsymbol{\alpha}\in \mathbb{R}^d\). To estimate the conditional survival function, when the outcome is continuous, we assume that the conditional probability density function follows a Gaussian model: 
\begin{equation*}
 f(y|\boldsymbol{X},A)=\frac{1}{\sqrt{2\pi\mathrm{exp}\{\xi_0-(\boldsymbol{X}^T, A, \boldsymbol{X}^T A)\boldsymbol{\xi}\}}}\mathrm{exp}\left[-\frac{\{y-\theta_0-(\boldsymbol{X}^T, A, \boldsymbol{X}^T A)\boldsymbol{\theta}\}^2}{2\mathrm{exp}\{\xi_0-(\boldsymbol{X}^T, A, \boldsymbol{X}^T A)\boldsymbol{\xi}\}}\right],
\end{equation*}
where the parameters \(\theta_0, \xi_0\in\mathbb{R}\) and \(\boldsymbol{\theta}, \boldsymbol{\xi}\in \mathbb{R}^{2d+1}\). 
The parameters can be estimated using the maximum likelihood method. 

For discrete outcomes, we focus on count outcomes in this paper. To model count outcomes, we assume the conditional probability mass function follows a negative binomial model: 
\begin{align*}
  p(y|\boldsymbol{X},A)=\left(\begin{array}{c}
    y+r_0+rA-1 \\
    y
    \end{array}\right)&\left[1-\mathrm{expit}\{\theta_0+(\boldsymbol{X}^T, A, \boldsymbol{X}^T A)\boldsymbol{\theta}\}\right]^{y} \\
    &\mathrm{expit}\{\theta_0+(\boldsymbol{X}^T, A, \boldsymbol{X}^T A)\boldsymbol{\theta}\}^{r_0+rA},\ y\in \mathbb{N},
\end{align*}
where the parameters \(r_0,r \in \mathbb{N}^+\), \(\theta_0\in\mathbb{R}\), and \(\boldsymbol{\theta}\in \mathbb{R}^{2d+1}\). The parameters can be estimated by \texttt{glm.nb} function in the R package \texttt{MASS}. For other types of discrete outcomes, an ordered probit model can be used for ordinal data, while a logistic model is appropriate for binary outcomes. 

\begin{remark}
We consider fitting parametric models when estimating these nuisance functions. In practice, if greater flexibility is needed, one could use more flexible machine learning models.
\end{remark}

\section{Theoretical property}  \label{sec:consistency}
\subsection{Result for continuous outcomes} \label{sec:resultcontinous}
{In this subsection, we aim to establish the theoretical properties of our estimator \(\hat{d}\), the output of Algorithm \ref{alg_1}, when the outcome is continuous.}
In particular, we derive the convergence rate of \( Q_\tau\{Y(\hat{d})\} - Q_\tau\{Y({d}^*)\} \) when the Gaussian kernel is used in constructing the estimated optimal decision functions \(\hat{f}_q\) in Algorithm \ref{alg_1}. 


Before presenting our main theorems, we need the following assumptions. 

\vspace{0.3ex}
\textbf{Assumption 2:} The counterfactual survival functions under treatment \(1\) and \(0\), \(S({q},1)\) and \(S({q},0)\), are continuous functions of \(q\).

\textbf{Assumption 3:} The nuisance models \(\hat{\pi}(a| \boldsymbol{x}), \hat{g}(q; \boldsymbol{x},a)\), \(q\in(l,r)\), as functions of \(\boldsymbol{x}\) and \(a\) are in the Vapnik–Chervonenkis (VC) classes \(\mathcal{F}_{\pi}\) and \(\mathcal{F}_{g}\), respectively.

\textbf{Assumption 4:} {There exists a constant \( \epsilon>0\)} such that 
\(P\{\epsilon<\hat{\pi}(a| \boldsymbol{X})<1-\epsilon\}=1 \) for \(a\in\{0,1\}\). 

\textbf{Assumption 5:} {For each fixed \((\boldsymbol{x},a)\),} \(\hat{\pi}(a|\boldsymbol{x})\) and \(\hat{g}(q;\boldsymbol{x},a)\) converge in probability to \(\pi^0(a|\boldsymbol{x})\) and \(g^0(q;\boldsymbol{x},a)\) uniformly in \(q\in(l,r)\) {with either \(\pi^0=\pi^*\) or \(g^0=g^*\)}, where the interval \((l,r)\) contains \((l^1,r^1)\) as defined in Algorithm \ref{alg_1}.  Moreover, {there exists a constant \(\gamma>0\)} such that,
\begin{align*}
    \Delta_n=\max_{a\in\{0,1\}}\sup_{q\in(l,r)}E\{|\psi(a,q;\hat{\pi},\hat{g})-\psi(a,q;\pi^0,g^0)|\}=O_p(n^{-\gamma}).
\end{align*}

These assumptions were used in the related literature when a classification method is employed to estimate the OTRs \cite[]{zhao2015doubly}. VC class (Assumption 3) includes regular parametric models (e.g., the conditional Gaussian model and the generalized linear model) and infinite-dimensional classes when certain smoothness or boundedness are satisfied \cite[]{van1996weak}. 
Assumption 4 is required to ensure the estimator is well-defined with nonzero denominators. {Assumption 5 implies that the estimated propensity score \(\hat{\pi}\) and the estimated conditional survival function \(\hat{g}\) converge to fixed functions \(\pi^0\) and \(g^0\), and it requires only one of them is correctly specified, i.e., either \(\pi^0=\pi^*\) or \(g^0=g^*\), suggesting double robustness of our method.} If we utilize parametric models to estimate \(\pi^*\) and \(g^*\), then \(\gamma=1/2\) in Assumption 5. 


To derive the convergence rate of \(Q_\tau\{Y(\hat{d})\}-Q_\tau\{Y({d}^*)\}\), we also require some regularity conditions on the distribution. 
Define \(\tau(\boldsymbol{x},q)=g^*(q; \boldsymbol{x},1)-g^*(q; \boldsymbol{x},0)\). Let
 \[
  \eta(\boldsymbol{x},q)=\frac{E\{{C}_{+}(\boldsymbol{X},q) | \boldsymbol{X}=\boldsymbol{x}, A=1\}-E\{{C}_{-}(\boldsymbol{X},q) | \boldsymbol{X}=\boldsymbol{x}, A=0\}}{E\{{C}_{+}(\boldsymbol{X},q) | \boldsymbol{X}=\boldsymbol{x}, A=1\}+E\{{C}_{-}(\boldsymbol{X},q) | \boldsymbol{X}=\boldsymbol{x}, A=0\}}+1 / 2,
 \]
 where \({C}_{+}(\boldsymbol{X},q)\) is an abbreviate of \(\max\{C(\boldsymbol{X},q;{\pi}^0,{g}^0),0\}\), \({C}_{-}(\boldsymbol{X},q)\) is an abbreviate of \\
 \(\max\{-C(\boldsymbol{X},q;{\pi}^0,{g}^0),0\}\), and \(C(\boldsymbol{X},q;{\pi}^0,{g}^0)\) is defined in (\ref{eq:itrs}). When either \({\pi}^0=\pi^*\) or \({g}^0=g^*\), \(E\{{C}_{+}(\boldsymbol{X},q)| \boldsymbol{X}=\boldsymbol{x}, A=1\}-E\{{C}_{-}(\boldsymbol{X},q) | \boldsymbol{X}=\boldsymbol{x}, A=0\}=\tau(\boldsymbol{x},q)\). Furthermore, we introduce the sets \( \mathcal{X}^{+}_q=\{\boldsymbol{x} \in \mathcal{X}: 2 \eta(\boldsymbol{x},q)-1>0\} \) and \( \mathcal{X}^{-}_q= \{\boldsymbol{x} \in \mathcal{X}: 2 \eta(\boldsymbol{x},q)-1<0\} \). A distance function to the boundary between \( \mathcal{X}^{+}_q\) and \( \mathcal{X}^{-}_q \) is denoted by \( \Delta(\boldsymbol{x},q)=\tilde{d}\left(\boldsymbol{x}, \mathcal{X}^{+}_q\right) \) if \( \boldsymbol{x} \) belongs to \( \mathcal{X}^{-}_q\), \( \Delta(\boldsymbol{x},q)= \tilde{d}\left(\boldsymbol{x}, \mathcal{X}^{-}_q\right) \) if \( \boldsymbol{x} \) lies within \( \mathcal{X}^{+}_q \), and \( \Delta(\boldsymbol{x},q)=0 \) otherwise, where \( \tilde{d}(\boldsymbol{x}, \mathcal{O}) \) represents the Euclidean norm-based distance of \( \boldsymbol{x} \) to set \( \mathcal{O} \). Recall that \({q}^*=\max_d Q_{\tau}\{Y({d})\}\) and \(d_q^*=\argmax_d S(q,d)\). Define, for \(\delta>0\),
\(
  Q_\delta=\{q: |S(q,d_q^*)-1+\tau|< \delta\}.
\)

The following assumptions are required.

\vspace{0.3ex}
\textbf{Assumption 6:} The distribution of \((\boldsymbol{X},A,Y)\) has geometric noise exponent \(\beta\) near \({q}^*\). Namely, there exist some positive constants \( C \), \(\delta^*\), and \(\beta\) such that
\begin{equation*}
  \sup_{q\in Q_{\delta^*}}E\left[\exp \left(-\frac{\Delta(\boldsymbol{X},q)^{2}}{t}\right)|2 \eta(\boldsymbol{X},q)-1|\right] \leq C t^{\beta d / 2}, t>0 .
\end{equation*}

\textbf{Assumption 7:} There exist some positive constants \(c\), \(\delta^*\), and \(\eta\) such that $  \sup_{q\in Q_{\delta^*}} P\{0<|\tau(\boldsymbol{X},q)|<t\}\leq ct^{\eta}, t>0 .$

\textbf{Assumption 8:}
There exist some positive constants \(\delta^*\), \(L_1\), and \(L_2\) such that, for any \(q,q_0\in Q_{\delta^*}\), $
   L_1|q_0-{q}^*|\leq |S(q_0,d_{q}^*)-S({q}^*,d_{q}^*)|\leq L_2|q_0-{q}^*|.$

In the related literature, assumptions similar to Assumptions 6--8 are typically stated at a fixed value \(q\) \cite[]{zhao2012estimating,zhou2017residual,shi2020sparse,ma2011asymptotic}. We generalize these assumptions to hold over a neighborhood \( Q_{\delta^*}\). Since \(\delta^*\) can be very small, our assumptions are only slightly stronger than those in the existing literature. We require such stronger assumptions to accommodate that the final classification to derive the estimated quantile OTR \(\hat{d}_{\hat{q}}\) relies on the estimated optimal quantile \(\hat{q}\) rather than a fixed \(q\).

Assumption 6 is introduced to guarantee the proper behavior of the classification procedure deriving \(\hat{f}_q\) in Algorithm \ref{alg_1}.
The geometric noise exponent \(\beta\) (Assumption 6) describes how fast the density of the distance \(\Delta(\boldsymbol{X},{q}^*)\) decays near the decision boundary, i.e., the \(\boldsymbol{x}\) satisfying \(2\eta(\boldsymbol{x},{q}^*)-1=0\). When the decay is rapid, there are fewer data points near the boundary, making classification easier, and allowing for the selection of a larger geometric noise exponent \(\beta\). An example is distinctly separable data, that is, when \(|2\eta(\boldsymbol{X},{q}^*)-1|>\delta_0>0\) with probability \(1\) for some constant \(\delta_0\), and \(\eta(\boldsymbol{x},{q})\) is continuous, \(\beta\) can be arbitrarily large. Assumption 7 is introduced to ensure the evaluation procedure in Algorithm \ref{alg_1} works properly, i.e., to bound the difference between \(\widehat{S}_{h_n}(q, \hat{f}_q)\) and \( S(q, d_q^*)\). When \(\tau(\boldsymbol{X},{q}^*)\) has a uniformly bounded density function near \(0\), and \(\tau(\boldsymbol{x},{q})\) is continuous, Assumption 7 holds with \(\eta=1\). If there exists some \(\delta_0>0\) such that \(|\tau(\boldsymbol{X},{q}^*)|> \delta_0\) with probability \(1\), and \(\tau(\boldsymbol{x},{q})\) is continuous, then Assumption 7 holds with an arbitrary large \(\eta\). In Assumption 8, for a given \(q\), when \(S(q_0, d_q^*)\) is differentiable at point \(q_0 = {q}^*\) with a nonzero derivative, i.e., the density function for \(Y(d_q^*)\) at \({q}^*\) is nonzero, both upper and lower bounds for \(|S(q_0, d_q^*) - S({q}^*, d_q^*)|\) in Assumption 8 can be established with a proper choice of \(\delta^*\). Thus, Assumption 8 imposes a weaker condition than requiring differentiability. 



For simplicity in notations, we also assume \(\lambda_n\), \(\sigma_n\), and \(h_n\) are the same across different \({q}\). 

Next, we present the main theorem, which establishes the convergence rate of our method using a Gaussian kernel.

\begin{theorem} \label{theorem3}
For continuous outcomes under Assumptions $1$--$8$, assume the following conditions hold as $n \to \infty$:
\(\lambda_n \to 0\), \(h_n \to 0\), \(\lambda_n n^{\min\{2\gamma,1\}} \to \infty\), and \(\lambda_n n h_n^2 \to \infty\).
Suppose further that
the initial interval \((l^1, r^1)\) for Algorithm \ref{alg_1} contains the optimal quantile $q^*$. Let $\zeta > 0$, $0 < v < 2$, \(\sigma_n = c_1 \lambda_n^{-1 / \{(\beta+1)d\}}\), \(\epsilon_n = c_2 n^{-1/2}\), and \(\kappa_n = c_3 n^{-1/2}\), where $c_1, c_2, c_3 > 0$ are constants. Then we have
\[
Q_\tau\{Y(\hat{d})\} - Q_\tau\{Y(d^*)\} = o_p(\varsigma_n \log n),
\]
where
  \(
  \varsigma _n=e^{-1 / (8h_n^2)}+\sqrt{c_n}/\sqrt{\lambda_nnh_n^2}+\lambda_n^{\beta\eta/\{(\beta+1)(\eta+1)\}}+\{c_n/(\lambda_nn)\}^{\eta/\{2(1+\eta)\}}+(\Delta_n^2/\lambda_n)^{\eta/\{2(1+\eta)\}}
  \)
and
$
c_n = \lambda_n^{-(2 - v)(1 + \zeta) / \{2(\beta+1)\}}.
$ 
\end{theorem}

Since the quantile OTRs may be non-unique \cite[]{zhouTransformationInvariantLearningOptimal2022}, Theorem \ref{theorem3} primarily focuses on the quantile value consistency, rather than the consistency of the parameters for the quantile OTRs themselves. As Assumption 5 holds when either the model for the propensity score \(\pi^*\) or the conditional survival function \(g^*\) is correctly specified, i.e., either \(\pi^0=\pi^*\) or \(g^0=g^*\), Theorem \ref{theorem3} reveals that the estimated quantile OTR \(\hat{d}\) of our method is doubly robust in the sense that the quantile given \(\hat{d}\) will be consistent with the optimal quantile. { The bound in Theorem \ref{theorem3} depends on the convergence rates of the estimated propensity score \(\hat{\pi}\) and the estimated conditional survival function \(\hat{g}\) toward their limits \(\pi^0\) and \(g^0\), as captured by \(\Delta_n\) defined in Assumption 5. In general, the same bound for \(\Delta_n\), and hence the same bound in Theorem~\ref{theorem3}, can be achieved even when only one of \(\pi^0=\pi^*\) and \(g^0=g^*\) holds. This is a distinctive asymptotic property for doubly robust classification, on which our method is built \cite[]{zhao2015doubly}. We refer the reader to Section 5.5 of the supplementary material for further details.
}

The bound in Theorem~\ref{theorem3} reveals a bias–variance trade-off governed by the choices of \(h_n\) and \(\lambda_n\). The bias term associated with \(h_n\), \(e^{-1 / (8h_n^2)}\), is induced by the difference between the kernel smoother \(\Phi\{\hat{f}_q(\boldsymbol{x})/h_n\}\) and \(I\{\hat{f}_q(\boldsymbol{x})>0\}\) for \(\tau(\boldsymbol{x},q)\neq 0\),
where \(\Phi(x),x\in\mathbb{R},\) is the cumulative distribution function of the standard normal distribution. The bias follows an exponential rate because 
\(|\Phi\{\hat{f}_q(\boldsymbol{x})/h_n\} - I\{\hat{f}_q(\boldsymbol{x}) > 0\}|\) exhibits behavior akin to the tail of the normal distribution for \(\tau(\boldsymbol{x},q)\neq 0\).
The bias term associated with \(\lambda_n\), \(\lambda_n^{\beta\eta/\{(\beta+1)(\eta+1)\}}\), is determined by the approximation properties of the Gaussian kernel. 

The optimal rate for the value function is \(\max\{ n^{-2\beta\eta /[\{6 \beta+2+(2-v)(1+\zeta)\}(\eta+1)]},n^{-2\gamma\beta\eta/\{(3\beta+1)(\eta+1)\}}\}\) up to a logarithm with \(h_n=1/\log n\) and \(
  \lambda_n=\max\{ n^{-2(\beta+1) /\{6 \beta+2+(2-v)(1+\zeta)\}},n^{-2\gamma(\beta+1)/(3\beta+1)}\}
\).
The rate consists of two parts: the first part reflects the rate of convergence in estimating the quantile OTRs; the second part is related to nuisance function estimation. When data are well separated, which implies that \(\beta\) and \(\eta\) in Assumptions 6 and 7 can be arbitrarily large, then the convergence rate almost achieves the rate \(\max\{n^{-1/3},n^{-2\gamma/3}\}\). When the nuisance models are estimated with convergence rates \(\Delta_n=O_p(n^{-1/2})\) such that \(\gamma=1/2\), which is satisfied for most parametric estimators, the rate for the value function is about \(n^{-1/3}\). 
Our result differs from those in \citet{zhao2015doubly,zhou2017residual}, as we focus on the convergence rate of the quantile difference, whereas the existing literature primarily investigates the difference in expectations, i.e., \(E\{Y(\hat{d})\}-\max_dE\{Y(d)\}\).

If we know the optimal quantile \(q^*\) as a priori, quantile OTRs can also be estimated by (\ref{surrogate}) with \(q=q^*\). The convergence rate for this oracle method is \(\max\{ n^{-2\beta /\{6 \beta+2+(2-v)(1+\zeta)\}},\)\\\(n^{-2\gamma\beta/(3\beta+1)}\}\) as Section S5.6 of the supplementary material suggests. When the nuisance functions are estimated with convergence rates \(\Delta_n=O_p(n^{-1/2})\) and the data are well separated, the convergence rate for this oracle method is about \(n^{-1/3}\). This indicates that the convergence rate of our method is close to the convergence rate of the oracle method in the well-separated case. 

\subsection{Result for discrete outcomes} \label{sec:resultdiscrete}
In this subsection, we establish theoretical guarantees for our method when the outcomes are discrete. As our approach is designed to maximize the smoothed quantile, 
we begin by demonstrating the convergence rate of \(Q^m_{\tau}\{Y(\hat{d})\}-Q^m_{\tau}\{Y(\tilde{d}^*)\}\) when a Gaussian kernel is used to construct \(\hat{d}\), where \(Q_{\tau}^m\) is the smoothed quantile and \(\tilde{d}^*=\argmax_d Q_{\tau}^m\{Y(d)\}\). {We then establish that \(Q_{\tau}\{Y(\hat{d})\}\) equals the optimal quantile \(\max_dQ_{\tau}\{Y({d})\}\) with probability tending to 1. Namely, \(\lim_{n\to\infty} P[Q_{\tau}\{Y(\hat{d})\}=\max_dQ_{\tau}\{Y({d})\}]=1\), a special asymptotic result that is typically observed and of interest in the context of discrete quantile function \cite[]{machado2005quantiles,chen2010quantile}.}
Finally, we show that our method is consistent in quantile value and effective under the homoscedastic treatment effect assumption. 

\vspace{0.3ex}
\textbf{Assumption 9:} (The homoscedastic treatment effect assumption.) The sign of \(P(Y>q| \boldsymbol{X}=\boldsymbol{x},A=1)-P(Y>q| \boldsymbol{X}=\boldsymbol{x},A=0)\) remains invariant for all \(q \in [v_1, v_l)\).

Assumption 9 holds for binary outcomes and count outcomes following the conditional Poisson distribution. See Section S5.2 of the supplementary material for details. 

Since the smoothing technique maps discrete outcomes to continuous ones, the proof strategy employed in Theorem \ref{theorem3} can be directly extended to establish the following convergence rate result for $Q^m_{\tau}\{Y(\hat{d})\} - Q^m_{\tau}\{Y(\tilde{d}^*)\}$, with Assumptions 2 and 6–8 appropriately modified to Assumptions $2^*$ and $6^*$–$8^*$. The modified assumptions are provided in Section S6.2 of the supplementary material. Then, using the convergence of \(Q^m_{\tau}\{Y(\hat{d})\}-Q^m_{\tau}\{Y(\tilde{d}^*)\}\), we can establish that \(Q_{\tau}\{Y(\hat{d})\}\) equals the optimal quantile \(\max_dQ_{\tau}\{Y({d})\}\) with probability tending to 1. These results are summarized in Theorem \ref{theorem4}.

\begin{theorem} \label{theorem4}
For discrete outcomes under Assumptions $1, 2^*$, $3-5$, $6^*-8^*$, assume the following conditions hold as $n \to \infty$:
\(\lambda_n \to 0\), \(h_n \to 0\), \(\lambda_n n^{\min\{2\gamma,1\}} \to \infty\), and \(\lambda_n n h_n^2 \to \infty\).
Suppose further that the initial interval \((l^1, r^1)\) contains the optimal smoothed quantile $\max_d Q_{\tau}^m\{Y(d)\}$. Let $\zeta > 0$, $0 < v < 2$, \(\sigma_n = c_1 \lambda_n^{-1 / \{(\beta+1)d\}}\), \(\epsilon_n = c_2 n^{-1/2}\), and \(\kappa_n = c_3 n^{-1/2}\), where $c_1, c_2, c_3 > 0$ are constants. Then we have

(i)
\(
Q_\tau^m\{Y(\hat{d})\} - Q_\tau^m\{Y(\tilde{d}^*)\} = o_p(\varsigma_n \log n),
\)
where
  \(
  \varsigma _n=e^{-1 / (8h_n^2)}+\sqrt{c_n}/\sqrt{\lambda_nnh_n^2}+\lambda_n^{\beta\eta/\{(\beta+1)(\eta+1)\}}+\{c_n/(\lambda_nn)\}^{\eta/\{2(1+\eta)\}}+(\Delta_n^2/\lambda_n)^{\eta/\{2(1+\eta)\}}
  \)
and
$
c_n = \lambda_n^{-(2 - v)(1 + \zeta) / \{2(\beta+1)\}}.$ 

(ii) 
\(
\lim_{n\to\infty} P[Q_{\tau}\{Y(\hat{d})\}=\max_dQ_{\tau}\{Y({d})\}]=1.
\)
\end{theorem}

The assumptions underpinning Theorem \ref{theorem4} closely mirror those required for Theorem \ref{theorem3}. Consequently, we refer the reader to Section \ref{sec:resultcontinous} for a more comprehensive discussion of these assumptions.
The part (i) of this theorem shows that the convergence rate of $Q_\tau^m\{Y(\hat{d})\} - Q_\tau^m\{Y(\tilde{d}^*)\}$ for a discrete outcome is the same as the convergence rate of $Q_\tau\{Y(\hat{d})\} - Q_\tau\{Y(d^*)\}$ for a continuous outcome, as established in Theorem \ref{theorem3}. Thus, a comprehensive discussion of the convergence rate can be found in Section \ref{sec:resultcontinous}. 

The part (ii) of Theorem \ref{theorem4} demonstrates that \(Q_{\tau}\{Y(\hat{d})\}=\max_dQ_{\tau}\{Y({d})\}\) with probability tending to 1. Specifically, when the outcome $Y(d)$ is discrete, the quantile $Q_{\tau}\{Y({d})\}=\sup\{q: S(q,{d})>1-\tau\}$ is also discrete. Therefore, if \(\hat{d}\) is a reliable estimator of the quantile OTRs with respect to the quantile value, \(Q_{\tau}\{Y(\hat{d})\}\) should equal \(\max_dQ_{\tau}\{Y({d})\}\) with probability tending to 1. Otherwise, \(Q_{\tau}\{Y(\hat{d})\}\) would deviate significantly from the optimal quantile.

Since the part (ii) of Theorem \ref{theorem4} indicates that for any \(\varepsilon>0\), $\lim_{n \to \infty}P[|Q_{\tau}\{Y(\hat{d})\} - \max_d Q_{\tau}\{Y(d)\}|<\varepsilon]\geq \lim_{n \to \infty}P[Q_{\tau}\{Y(\hat{d})\} = \max_d Q_{\tau}\{Y(d)\}] = 1$, our method is quantile value consistency.
Next, we state that our method is effective under the homoscedastic treatment effects assumption:
\begin{corollary}
\label{pro:effective}
Under the assumptions of Theorem \ref{theorem4}, Assumption 9, and the boundedness condition $|Y| \leq M_y$, we have
\begin{align*}
E\{Y(\hat{d})\} - \max_d E\{Y(d)\} = o_p(1).
\end{align*}
\end{corollary}

This corollary demonstrates that the expected counterfactual outcome given the estimated regime approaches optimal expected outcomes under {Assumption 9}. Thus, our method is effective in this setting. While deriving explicit convergence rates for the expectation differences is possible with additional regularity conditions, such analysis extends beyond the scope of this paper.

\begin{remark}
The techniques employed to establish the part (ii) of Theorem \ref{theorem4} and Corollary \ref{pro:effective} can likewise be applied to other methods that incorporate our smoothing approach, thereby demonstrating that those methods are consistent in quantile value and effective by combining our smoothing technique. Indeed, as shown in Sections S6.2 and S6.3 of the supplementary material, if a method satisfies smoothed quantile value consistency, it ensures quantile value consistency. Additionally, if the data satisfies the homoscedastic treatment effect assumption and boundedness \(|Y|<M_y\), the method satisfying smoothed quantile value consistency is also effective.
\end{remark}

\section{Simulation} \label{simulations}
We evaluate our method through simulations. The baseline covariates \(X_1\) and \(X_2\) are independently drawn from a standard normal distribution. The treatment \( A \) is generated from a logistic regression model
\(
  \pi^*(A=1 | \boldsymbol{X})=\operatorname{expit}\left\{-0.5-0.5 (X_1+X_2)\right\}.
\)


We consider quantiles \(\tau=0.25\) and \(0.50\). We consider the following three different cases: 

\textbf{Case 1 (Continuous outcome with linear regime):} The continuous outcome \(Y\) is generated by \(Y=2-2X_1+2X_2+A(X_1+X_2)+(2A+1)\epsilon\), where \(\epsilon\) follows a standard normal distribution. In this case, the \(\tau\)-quantile OTRs are \(d^*(\boldsymbol{X})=I\{2{q}^*-4+5X_1-3X_2 > 0\}\), where \({q}^*=0.485\) for \(\tau=0.25\) and \({q}^*=2.789\) for \(\tau=0.50\). 

\textbf{Case 2 (Continuous outcome with nonlinear regime):} The continuous outcome \(Y\) is generated by \(Y=2-0.5X_1+0.5X_2^2+A(X_1+X_2)+\mathrm{exp}\{-2-0.5X_1-0.5X_2+ A(2+X_1+X_2)\}\epsilon\), where \(\epsilon\) follows a standard normal distribution. In this case, the quantile OTRs are nonlinear. We plot the boundaries of quantile OTRs for \(\tau=0.25\) and \(\tau=0.50\) in Figures S4 and S5 of the supplementary material.

\textbf{Case 3 (Discrete outcome with nonlinear regime):} The discrete outcome \(Y\) follows the negative binomial distribution with number of success \(2A+1\) and success probability \(\mathrm{expit}\{-2.5-0.5X_1+0.25X_2^2+A(1+X_1+X_2^2)\}\). Figures S6 and S7 in the supplementary material  illustrate the boundaries for quantile OTRs, which suggests that the quantile OTRs are nonlinear. 

We implement the following methods for estimating quantile OTRs:

1. SCL-Linear: Our proposed method with a linear kernel, with nuisance functions estimated as described in Section \ref{nuisance}.

2. SCL-Gaussian: Our proposed method with a Gaussian kernel, where the nuisance functions are estimated following the approach detailed in Section \ref{nuisance}.

3. Wang’s method \cite[]{wang2018quantileoptimal}: A doubly robust approach that leverages both the propensity score and conditional quantile information to ensure robustness. The propensity score is estimated using logistic regression models, as outlined in Section \ref{nuisance}, while the conditional quantile is estimated via quantile regression models incorporating the basis function set \((\boldsymbol{X}, A, \boldsymbol{X}A)\). Since the method is not directly applicable to discrete outcomes, we implement our smoothing technique for the estimated quantile function and then apply the genetic algorithm to maximize the estimated smoothed quantile to derive the quantile OTRs. See Section S2.2 of the supplementary material.

4. QIQ-Learning \cite[]{linn2017interactive}: A quantile interactive Q-learning approach that estimates the conditional survival function \( g^*(q;\boldsymbol{x}, a) \) and transforms it into quantile OTRs. The conditional survival function is estimated according to the procedure detailed in Section \ref{nuisance}. Since the method is not directly applicable to discrete outcomes, we implement our smoothing technique for the estimated conditional survival function and then transform the estimated conditional smoothed survival function into quantile OTRs. See Section S2.2 of the supplementary material.



\vspace{1ex}

The simulations are conducted with a sample size of 500, repeated 200 times, and assessed using a large independent test set of 100,000 subjects. In this study, we utilize two performance indicators: the value function given the estimated quantile OTRs
and the misclassification rate (MR) by comparing the true and the estimated quantile OTRs in test samples. For Cases 1 and 2, the value function is given by the quantile \( Q_{\tau} \), while for Case 3, it is defined as the smoothed quantile \( Q_{\tau}^m \).
The results are reported in Table \ref{simulation}. 

 \begin{table}
    \caption{Mean (with standard deviation in the parenthesis) of value functions (``Value'') and misclassification rates (``MR'') evaluated on independent test data for three cases. {The best result in each scenario is highlighted in bold.}} 
    \label{simulation}
    \centering
    \begin{threeparttable}
      \resizebox{0.99\columnwidth}{!}{
        \begin{tabular}{lcrrr}
          \toprule Method & Value for \(\tau=0.25\)  &   MR for \(\tau=0.25\)  & Value for \(\tau=0.50\)   & MR for \(\tau=0.50\)             \\
          \midrule 
          &\multicolumn{4}{c}{Case 1 (optimal values: 0.485 for \(\tau=0.25\) and 2.789 for \(\tau=0.50\))}  \\
           \cmidrule{2-5}
        SCL-Linear &$0.383$ ($0.141$) & $0.107$ ($0.078$) & $2.725$ ($0.080$) & $0.102$ ($0.067$) \\
    SCL-Gaussian&$0.384$ ($0.102$) & $0.117$ ($0.061$) & $2.702$ ($0.076$) & $0.123$ ($0.058$) \\
     Wang's method&$0.257$ ($0.229$) & $0.137$ ($0.066$) & $2.615$ ($0.091$) & $0.194$ ($0.089$) \\
    QIQ-learning&$\boldsymbol{0.477}$ ($0.017$) & $\boldsymbol{0.033}$ ($0.019$) & $\boldsymbol{2.783}$ ($0.013$) & $\boldsymbol{0.035}$ ($0.019$) \\

            \midrule 
          &\multicolumn{4}{c}{Case 2 (optimal values: 2.250 for \(\tau=0.25\) and 2.783 for \(\tau=0.50\))}  \\
          
        \cmidrule{2-5}
     SCL-Linear&$2.153$ ($0.022$) & $0.219$ ($0.021$) & $2.714$ ($0.017$) & $0.207$ ($0.037$) \\
    SCL-Gaussian&$\boldsymbol{2.211}$ ($0.019$) & $\boldsymbol{0.099}$ ($0.040$) & $\boldsymbol{2.743}$ ($0.016$) & $\boldsymbol{0.125}$ ($0.038$) \\
     Wang's method&$2.074$ ($0.028$) & $0.300$ ($0.041$) & $2.537$ ($0.020$) & $0.440$ ($0.027$) \\
    QIQ-learning&$2.150$ ($0.018$) & $0.198$ ($0.017$) & $2.688$ ($0.020$) & $0.204$ ($0.024$) \\
        \midrule 
          &\multicolumn{4}{c}{Case 3 (optimal values: 3.349 for \(\tau=0.25\) and 8.853 for \(\tau=0.50\))}  \\
       
        \cmidrule{2-5}
   SCL-Linear &$2.286$ ($0.214$) & $0.314$ ($0.053$) & $8.009$ ($0.184$) & $0.228$ ($0.011$) \\
   SCL-Gaussian&$\boldsymbol{3.018}$ ($0.244$) & $\boldsymbol{0.139}$ ($0.063$) & $\boldsymbol{8.479}$ ($0.223$) & $\boldsymbol{0.127}$ ($0.047$) \\
    Wang's method&$1.974$ ($0.206$) & $0.290$ ($0.048$) & $7.194$ ($0.354$) & $0.329$ ($0.051$) \\
   QIQ-learning&$2.146$ ($0.142$) & $0.314$ ($0.022$) & $7.693$ ($0.225$) & $0.253$ ($0.018$) \\

          \bottomrule
        \end{tabular}
      }
    \end{threeparttable}
  \end{table}

In Case 1, {QIQ-learning performs best, while our method shows satisfactory performance in value functions and MRs. This result is expected, as QIQ-learning relies solely on the outcome model; under a correctly specified conditional survival function, it outperforms our doubly robust estimator, which depends on both outcome and propensity score models. Wang’s method, specifically designed for linear quantile OTRs, performs less satisfactorily, likely due to some solutions being trapped in local minima.} In Case 2, the quantile OTRs are nonlinear. 
Our method utilizing the Gaussian kernel outperforms all other methods. This suggests that our approach with the Gaussian kernel is capable of adapting to nonlinear scenarios. 
Case 3 focuses on nonlinear quantile OTRs with discrete outcomes. Our method with the Gaussian kernel outperforms the alternatives, demonstrating its ability in addressing both the challenges posed by discrete outcomes and the nonlinearity of the quantile OTRs.

{In the supplementary material Section S3, we visualize the relationship between the estimated and true quantile OTRs. We also present additional simulation studies that consider a larger number of covariates and a range of sample sizes. Additional results, including computational time, the bias of the estimated optimal value function, and the estimated value function evaluated under the estimated quantile OTRs, are reported therein. Notably, our proposed approaches, SCL-Linear and SCL-Gaussian, demonstrate substantially greater computational efficiency than Wang’s method. In addition, we include simulation results in the supplementary material to validate the doubly robust property of our method under misspecification of the propensity score \(\pi^*\) or the conditional survival function \(g^*\),  for both the linear kernel and the Gaussian kernel.}

\section{Real data analysis} \label{realdata}
We illustrate our method on the ACTG175 dataset from the R package \texttt{speff2trial}, which contains measurements on 2139 HIV-infected patients. The patients were randomized to four treatment arms: zidovudine (AZT) monotherapy, AZT+didanosine (ddI), AZT+zalcitabine (ddC), and ddI monotherapy. The outcome we consider is the CD4 cell count (per cubic millimeter) at 20 ± 5 weeks post the baseline. A larger value indicates a better outcome. 
Same as \cite{wang2018quantileoptimal}, we treat the CD4 cell count as a continuous outcome. We focus on assigning treatments, specifically the AZT+ddI combination therapy or the ddI monotherapy, to patients, as \cite{wang2018quantileoptimal} revealed these two treatments exhibit heteroscedastic treatment effects.

We consider all twelve variables measured at baseline as the covariates for each subject followed by \cite{fan2017concordanceassisted}. Five of the twelve covariates are continuous or count variables: age (years), weight (kilogram), Karnofsky score (on a scale of 0-100), CD4 cell counts at baseline (per cubic millimeter), and CD8 cell counts at baseline (per cubic millimeter). The rest seven are binary: hemophilia (0=no, 1=yes), homosexual activity (0=no, 1=yes), history of intravenous drug use (0=no, 1=yes), race (0=white, 1=non-white), gender (0=female, 1=male), antiretroviral history (0=naive, 1=experienced), and symptomatic indicator (0=asymptomatic, 1=symptomatic).

We consider \(\tau = 0.25, 0.50\), and \(0.75\). The performance of our method is compared with other approaches discussed in Section \ref{simulations} using repeated random sample splitting.  {Specifically, we randomly split the dataset into training set and test set 200 times, allocating 80\% of the observations to the training set and the remaining 20\% observations to the test set.}  For each split, we estimate the quantile OTRs using each method based on the training set. 
We then evaluate the performance of the estimated quantile OTRs for each method by computing the estimated quantile \(\widehat{Q}_\tau\{Y(\hat{d})\}\) using the test set, where \(\widehat{Q}_\tau\{Y(\hat{d})\}\) is calculated as in (\ref{wang.quantile}).
%
The average of \(\widehat{Q}_\tau\{Y(\hat{d})\}\) across 200 splits are shown in Table \ref{real_data}. It is evident that our method, SCL-Gaussian, attains the highest quantile values at all the quantile levels, outperforming the competing appraches.

 \begin{table}
    \caption{Mean (standard deviation) of value functions (``Value'') for different methods. The best result in each scenario is highlighted in bold.} \label{real_data}
    \centering
  \begin{tabular}{lccc}
  \hline & Value for \(\tau=0.25\) & Value for \(\tau=0.50\) & Value for \(\tau=0.75\) \\
  \hline SCL-Linear & \( 281.8(4.4) \)  & \( {380.2}(8.5) \) &\( 496.4 (11.3)\) \\
  SCL-Gaussian & \( \boldsymbol{283.3}(4.8) \)  & \( \boldsymbol{383.2}( 8.3) \) & \(\boldsymbol{500.2} (9.8)\) \\
  Wang's method & \( 276.9(5.9) \)  & \( 368.1(9.8) \) & \(478.3(10.8)\)\\
  QIQ-learning & \( 280.3(4.4) \)  & \( 375.9(8.0) \) & \(490.6(8.3)\)\\
  \hline
  \end{tabular}
\end{table}

{To strengthen our finding, we also add {another} metric: the Rand index (RI) \cite[]{rand1971objective}, which quantifies the stability of different methods in assigning quantile optimal treatments. 
Specifically, we randomly divide the data into two subsets, the training and test sets, allocating 80\% of the observations to the former and the remaining 20\% to the latter.
Within the training set, we further partition the data into two equal halves and obtain two estimated quantile OTRs from each half, denoted by \(\hat{d}_1\) and \(\hat{d}_2\), respectively. We then compute the Rand index on the test set as \(\sum_{j=1}^m I\{\hat{d}_1(\boldsymbol{X}_j)=\hat{d}_2(\boldsymbol{X}_j)\}/m\). The closer the Rand index is to 1, the greater the stability of the method. Repeating this process 200 times, we report the average Rand index for different methods in Table \ref{real_data_RI_sub}. It is evident that our method with a Gaussian kernel achieves the highest stability in assigning quantile optimal treatments, as reflected by a larger Rand index relative to competing approaches.}

\begin{table}
    \caption{Mean (standard deviation) of Rand index (``RI'') for different methods. The best result in each scenario is highlighted in bold.} \label{real_data_RI_sub}
    \centering
  \begin{tabular}{lccc}
  \hline & RI for \(\tau=0.25\) & RI for \(\tau=0.50\) & RI for \(\tau=0.75\) \\
  \hline SCL-Linear &  \(0.657(0.223)\)   & \(0.823(0.202)\)  & \(0.813(0.195)\) \\
  SCL-Gaussian  &  \(\boldsymbol{0.801}(0.162)\)  &\(\boldsymbol{0.881}(0.151)\) & \(\boldsymbol{0.870}(0.178)\) \\
  Wang's method  & \(0.523(0.109)\) &\(0.554(0.108)\) & \(0.605(0.166)\)\\
  QIQ-learning  & \({0.614}(0.074)\)  & \(0.645(0.068)\) &\(0.643(0.073)\)\\
  \hline
  \end{tabular}
\end{table}

\section{Discussion} \label{sec:discussion}
In this paper, we introduce a successive classification learning method for estimating quantile OTRs. Our approach leverages the benefits of classification learning, both in optimization and statistical properties, to overcome the challenges posed by non-convexity and regime linearity inherent in previous methods. To tackle the inconsistency and ineffectiveness issues associated with discrete outcomes, we propose to smooth the quantile. Our method demonstrates satisfactory performance in both simulation and real data. {We further extend our method to longitudinal studies and survival data in Section S4 of the supplementary material to enhance the applicability of our framework.} 

This paper focuses on a binary treatment scenario. In certain diseases, multiple treatment options may be available. It would be of interest to extend our method to multiple treatments. We can adopt the idea from \cite{zhang2020multicategory} to embed all possible treatments into a series of points in Euclidean space and combine the angle-based method with Algorithm \ref{alg_1} to learn the quantile OTRs. The angle-based method is similar to classification learning, which utilizes machine learning techniques to enhance computational efficiency and flexibly accommodate nonlinear regimes.

In various applications, the baseline covariates may be large, and variable selection is important to have an efficient and explanatory estimation. We can incorporate the LASSO or SCAD penalty \cite[]{tibshirani1996regression,fan2001variable} to choose variables for the linear quantile OTRs. If we aim to derive a sparse nonlinear regime with the Gaussian kernel, the gradient-based variable selection method \cite[]{yang2016modelfree} can be employed.

\section*{Data availability statement}
The ACTG175 dataset that supports the findings of this study is openly available in the R package \texttt{speff2trial} at \url{https://cran.r-project.org/web/packages/speff2trial}.

\bibliography{An_anti-confounding_method}
\end{document}